\documentclass[12pt]{report}
\usepackage{amsthm,amssymb} 
\usepackage[pdftex]{graphicx}
\DeclareGraphicsExtensions{.pdf,.jpeg,.png}
\usepackage{enumerate}
\usepackage{epsfig}   
\usepackage{hyperref}
\usepackage[onehalfspacing]{setspace}
\usepackage{amsmath,bm}
\usepackage{fixmath}
\usepackage{amssymb}
\usepackage{textcomp}
\usepackage{subfig}
\usepackage{gensymb}
\usepackage{textcomp}

\renewcommand{\bibname}{References}

\renewcommand{\textwidth}{6in}

\begin{document}
\pagenumbering{roman}
\begin{singlespace}
%
%

\thispagestyle{empty}
\begin{center}
    \vspace{1cm}
    {\LARGE Finite State Markov Modeling of Fading Channels Towards Decoding of LDPC Codes}\\
    \vspace{.2cm}
    {A dissertation submitted in the partial fulfillment of the requirements for the degree of} \\
     \vspace{0.2cm}
     {\bf Master of Technology}\\
\vspace{0.2cm}
     { \emph{in}} \\
     \vspace{.2cm}
    {\bf Communication Engineering}\\

    \vspace{.8cm}
    { Submitted By}\\
    \vspace{0.2cm}
    {\large \bf Mohit Kumar} \\
\vspace{0.15cm}
{\bf (2008EEE3223)} \\
\vspace{0.8cm}
{ Under the Guidance of}\\
\vspace{0.2cm}
{\large \bf Prof. S. D. Joshi} \\
\vspace{1.4cm}
   \begin{figure}[th]
    \centering
        \scalebox{.9}{\includegraphics{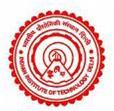}}

    \end{figure}\vspace{0.2cm}
    {\large \bf Department of Electrical Engineering}\\
    \vspace{0.2cm}
        {\large \bf Indian Institute of Technology, Delhi}\\
        \vspace{0.2cm}
       {\large \bf May, 2010}

\end{center}

\newpage
%
%
%
%
%
%

 \begin{center}
 {\LARGE \bf Certificate}
 \end{center}
 \vspace{1cm}
 This is to certify that the work contained in the dissertation titled {\textbf{``Finite State Markov Modeling of Fading Channels Towards Decoding of LDPC Codes"}} by \textbf{Mohit Kumar (2008EEE3223)} in partial fulfillment of the course work requirement of M.Tech program in the Department of Electrical Engineering has been carried out under my supervision. This work has not been submitted anywhere else for award of any degree or diploma.\\~\\~\\~\\

\hfill{ \textbf{Prof. S.D.Joshi}\\
Date:\\
Place: IIT Delhi}

\addcontentsline{toc}{chapter}{Certificate}
\newpage
%
%

\begin{center}
{\LARGE \bf Abstract}
\end{center}
\vspace{1.5cm}
Here we have proposed two decoding strategies of low-density parity check (LDPC) codes over Markov noise channels with bit flipping noise. The sum-product algorithm used for decoding LDPC codes over memoryless channels is extended to include channel estimation and how much gain we obtain by doing so is simulated and verified.
LDPC codes have been studied for years over memoryless channels and are known to have excellent performance. However, these codes over channels with memory is the topic of current research. Here, channels with memory are characterized by Markov modeling which is a useful busty channel model. With sufficient no. of states, they are able to model sufficient noise characteristics. We have gone for a two state system as it shows a good compromise between complexity and performance.


\addcontentsline{toc}{chapter}{Abstract}
\newpage
%
%
%
%

%
\begin{center}
{\LARGE \bf Acknowlegdement}
\end{center}
\vspace{1.5cm}
I would like to express my deep sense of gratitude to my project supervisor, \textbf{Prof. S. D. Joshi} for
his invaluable guidance, suggestions and constant encouragement without which this project would have been a dream unconquered. His experience and vast knowledge has
always proved to be a great asset in the thesis work. He has always
showed immense patience in solving my doubts and has channelized me
in right direction. I really appreciate his dedication for imparting
knowledge. I would also like to thank him for the
special lectures he offered during the evening session. He worked on our fundamentals and these discussions always bring up new ideas and give good
understanding of the various untouched topics.

I would also like to whole-heartedly thank my family and friends for always being very
supportive and cooperative.

\vspace{2cm}
\hfill{ \textbf{Mohit Kumar}}
%
%
%


%

\addcontentsline{toc}{chapter}{Acknowledgement}
\newpage
\tableofcontents
   \newpage
\listoftables
\addcontentsline{toc}{chapter}{List of Tables}
\newpage
\listoffigures
\addcontentsline{toc}{chapter}{List of Figures}   %
\end{singlespace}
\newpage
\pagenumbering{arabic}
%
%
%

%
\vspace{1.5cm}

\chapter{Introduction}
\begin{figure}
\begin{center}
  \scalebox{0.8}{\includegraphics{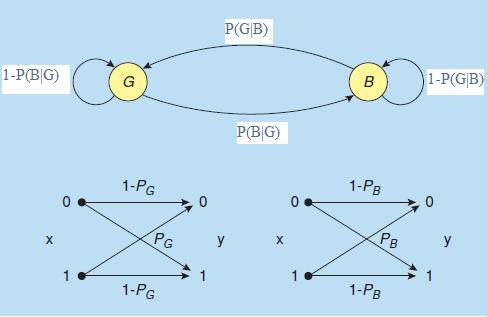}}

  \caption{GEC Model for a flat fading channel}\label{fig:fig30}
\end{center}
\end{figure}

Burst noise modeling and decoding at the receiver incorporating Finite State Markov Modeling (FSMC) and Gilbert-Elliott channel (GEC) are widespread for a variety of applications now a days. Also recent research in error correcting codes have led us to Low Density Parity Check codes (LDPC) with reasonable decoding complexity able to achieve reliable transmission close to the Shannon limit. Gallager \cite{a1} first presented these codes and demonstrated its performance for a memoryless Binary Symmetric Channel (BSC). This thesis extends the development of FSMC models and how these models can aid in the development of efficient algorithms at the receiver end giving a great performance advantage.\\
The study of finite-state communication channels with memory
dates back to the work by Shannon in 1957. In 1960, Gilbert introduced a new type of finite-state channel (FSC) model to
determine the information capacity of wireline telephone circuits with burst-noise. Gilbert model was the first attempt to incorporate channel memory into the
system with different states, which are unknown to the receiver. Soon after the work of Gilbert , Elliott started to compare performances of various error correcting codes using this model. This finally came to be known as GEC, as shown in figure \ref{fig:fig30}.

It is a two state model with states labeled as good G and the bad B state. The transition probabilities are given by $P(B|G) = Pr(G \rightarrow B)$ which gives the transition probability from Good to Bad state and $P(G|B) = Pr(B \rightarrow G)$ which gives the transition probability from Bad to Good state. When the channel is in good state G, the input and the output are related by a discrete memoryless channel (DMC) characterized by error probability $Pr(error \mid state=G)$. Similarly for the bad state, the error probability is given by $Pr(error \mid state=B)$.

The credit of inventing Low Density Parity check codes (LDPC) codes goes to Gallager \cite{a1}. He performed the detailed performance analysis of these codes. An LDPC code is a linear block code specified by a very sparse parity check matrix (H). An LDPC code is represented by a bipartite graph with $N$ symbol nodes, $M$ check nodes and edges connecting these symbol nodes to check nodes, if the $H$ matrix of the LDPC codes has $N$ rows and $M$ columns. There is an edge connecting the two nodes if there is a 1 existing in the corresponding entry in the $H$ matrix.\\
 We can categorize LDPC codes into regular and irregular LDPC codes. Regular codes have all nodes of the same degree. A $(n,j,k)$ regular LDPC code has a bipartite graph in which all symbol nodes have degree $j$ and check nodes of degree $k$, of length $n$. For irregular LDPC codes , the symbol nodes (correspondingly the check nodes) can have varying degrees. MacKay has presented extensive simulation results of long LDPC codes with the sum-product algorithm.

\section{Motivation}
Errors encountered in digital transmission over most real communication channels are not independent but appears in clusters. Such channels are said to exhibit memory and thus cannot be adequately represented by classical memoryless models. The existence of memory means additional capacity. To exploit this efficiently motivates us to effectively model the channel characteristics and use in Low Density Parity Check codes. A detailed study of channels with memory can be obtained in \cite{a14}.\\
In this thesis, we have proposed an iterative decoding algorithms for Low Density Parity Check codes(LDPC) for markov noise channels. The markov noise channel is an additive noise channel whose noise statistics is modeled by a markov model. The conventional sum-product algorithm is unable to fully extract the burst error correcting capability of LDPC codes. The proposed algorithms effectively utilizes the error correlation and uses it effectively to correct errors far beyond the random error correcting capability of the code. Further, it has been discovered recently that the combination of long LDPC codes together with sum-product decoder can provide near capacity performance over binary symmetric channels. My work extends it to channels with memory. Excellent performances of these codes based on very sparse matrices has encouraged researchers to have a look into the channels with memory, apart from the memoryless case.

\section{Thesis Organisation}
In chap 2, we provide the fundamentals for understanding the bipartite graph over which our decoding strategies are based. The sum-product algorithm is also introduced there. Chap 3 describes a new iterative decoding technique using a sum-product decoder and a helper block using the state information to aid in the decoding task. This chapter also gives a method for estimating the markov modeled channel parameters (error probabilities and transition probabilities). Chap 4 gives another algorithm for decoding LDPC codes on channels with memory and also gives a notion of decoding region where successful decoding can take place. Chap 5 gives an error predictor which can also aid the sum-product decoder. It is only introduced and not realized or verified through simulation. The next chap presents the various simulation results and the conditions assumed for carrying out such simulations. A brief description of various results is presented there. Finally we conclude our thesis with conclusion and future scope of work.

\chapter{An Introduction to Factor Graphs and Sum-Product Algorithm}
Various iterative decoding techniques have become a good alternatives for decoding systems. Most of these decoding techniques for codes utilize factor graphs or tanner graphs (as known previously). The main motto of factor graphs is to describe codes by means of equation systems, whose structure are the basis for decoding algorithms. The equation system defines a bipartite graph with vertices both for the variables and for the equations; an edge indicates that a particular
variable is present in a particular equation.
Factor graph is a way of realizing complicated global functions which render themselves to be factored as smaller factors comprising of a subset of variables of the global function. The sum product algorithm computes the various marginal functions associated with the
global function. These factor graphs are generalizations of Tanner graphs which Tanner had proposed for analyzing LDPC codes of Gallager. In the work of Tanner, all the nodes were visible, whereas Wiberg came up with \emph{hidden} nodes as well.
\section{Marginal Functions and factor graphs}
Let $x_{1},x_{2}...x_{n}$ be a collection of variables, in which,
for each $x_{i}$, takes on values in some domain $A_{i}$ . Let $g(x_{1},...,x_{n})$ be a function
of these variables.
Associated with every function $g(x_{1},...,x_{n})$, are $n$ marginal functions $g_{i}(x_{i})$. For each $a \in A_{i}$, the value of $g_{i}(a)$ is obtained by
summing the value of $g(x_{1},...,x_{n})$ over all configurations of
the variables that have $x_{i}=a$. This type of summation is so common here that we introduce a summary operator. Instead of indicating which variables are being summed over, we indicate the ones \emph{not} being summed over. This notation is taken from \cite{a15}. If $h$ is a function of three variables, $x_{1},x_{2}$ and $x_{3}$, then the \emph{summary} for $x_{2}$ can be denoted as
\begin{equation}
\sum_{\sim {x_{2}}} h(x_{1},x_{2},x_{3}) = \sum_{x_{1} \in A_{1}} \sum_{x_{3} \in A_{3}} h(x_{1},x_{2},x_{3})
\end{equation}

Thus,
\begin{equation}
g_{i}(x_{i}) = \sum_{\sim x_{i}} g(x_{1},...,x_{n})
\end{equation}

Suppose that $g(x_{1},...,x_{n})$ factors into a product of several
\emph{local functions}, each having some subset of $\{x_{1},...x_{n}\}$ as
arguments, then a factor graph is a bipartite graph that expresses
the structure of the factorization. A factor graph has a variable
node for each variable $x_{i}$ , a \emph{factor node} for each local function $f_{j}$
, and an edge-connecting variable node $x_{i}$ to factor node $f_{j}$
if and only if $x_{i}$ is an argument of $f_{j}$.

\section{\emph{Computing Marginal functions}}
What we are interested in computing $g_{i}(x_{i})$
for more than one value of $i$ or for all $i$'s. This can be achieved by an algorithm known as sum-product algorithm. It works by computing various sums and products at nodes.\\
The sum-product algorithm has the following simple rule: The message sent from a node $v$ on an edge $e$ is the
product of the local function at $v$ with all messages received at $v$
on edges \emph{other} than $e$, summarized for the variable
associated with $e$.\\
Let $\mu_{x \rightarrow f}(x)$ denote the message sent from $x$ node to $f$ node
in the operation of the sum-product algorithm, let $\mu_{f \rightarrow x}(x)$
denote the message sent from node $f$ to node $x$. Also, let $n(v)$
denote the set of neighbors of a given node $v$ in a factor graph. Then the message computations may be expressed as follows:\\
\textbf{\emph{Variable to local function:}}
\begin{equation}
\mu_{x \rightarrow f}(x) = \prod_{h \in n(x)\backslash \{f\}} \mu_{h \rightarrow x}(x)
\end{equation}\\
\textbf{\emph{Local function to variable:}}
\begin{equation}
\mu_{f \rightarrow x}(x) = \sum_{\sim x}\left(f(X) \prod_{y \in n(f) \backslash \{x\}} \mu_{y \rightarrow f}(y)\right)
\end{equation}
where $X=n(f)$ is the set of arguments of the function $f$. These messages are computed and sent along the edges to corresponding nodes. Once a node receives all the messages along the edges it is connected to, the computations takes place and a new message is formed. This is known as \emph{message passing schedule}.\\
Thus we can observe that variable nodes of degree two has to perform
no computation. A message arriving on one (incoming) edge is
simply transferred to the other (outgoing) edge.
\section{Probabilistic Modeling}
\begin{figure}
\begin{center}
  \scalebox{0.8}{\includegraphics{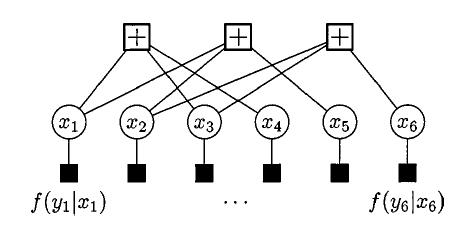}}

  \caption{Factor graph for the joint APP distribution of codeword symbols}\label{fig:fig2}
\end{center}
\end{figure}

Here we consider the standard
coding model where a codeword $x=\{x_{1},...,x_{n}\}$ is
selected from a code $C$ of length $n$ and transmitted over a
memoryless channel with corresponding output sequence $y=\{y_{1},...,y_{n}\}$. Thus for each observation $y$ , the joint
a posteriori probability (APP) distribution of $x$ (i.e., $p(x|y)$ ) is \emph{proportional} to the function $g(x)=p(y|x)p(x)$, where $p(x)$ is the a \emph{priori} distribution
for the transmitted vectors, and $p(y|x)$ is the conditional
probability density function for $y$ when $x$ is transmitted.\\
Having assuming a \emph{priori} distribution for the transmitted
vectors to be uniform over codewords, we have $p(x)=\chi_{c}(x)/|C|$,
where $\chi_{c}(x)$ is the characteristic function for $C$ and $|C|$ is the
number of codewords in $C$. If the channel is memoryless, then $p(\textbf{y}|\textbf{x})$
factors as
\begin{equation}
p(y_{1},...,y_{n}|x_{1},...,x_{n}) = \prod_{i=1}^{n} p(y_{i}|x_{i})
\end{equation}
Thus, we have,
\begin{equation}\label{e:eq2}
g(x_{1},...,x_{n}) = \frac{1}{|C|} \chi_{c}(x_{1},...,x_{n}) \prod_{i=1}^{n}p(y_{i}|x_{i})
\end{equation}
If we are given a factor graph $F$ for $\chi_{c}(x)$ , we obtain a
factor graph for (a scaled version of) the APP distribution over $x$
simply by \emph{augmenting} $F$ with factor nodes corresponding to the
different factors $p(y_{i}|x_{i})$. The $i^{th}$ such factor has only
one argument, namely $x_{i}$, since $y_{i}$ is regarded as a parameter.
Thus, the corresponding factor nodes appear as pendant vertices
(dongles) in the factor graph.\\
For example, if $C$ is a binary linear code with the check equation set as given below,we have,
\begin{equation}
g(x_{1},...,x_{6}) = [x_{1} \oplus x_{2} \oplus x_{5} := 0].[x_{2} \oplus x_{6} \oplus x_{5} := 0].[x_{1} \oplus x_{3} \oplus x_{4} := 0].\prod_{i=1}^{6} p(y_{i}|x_{i})
\end{equation}
whose factor graph is shown in figure \ref{fig:fig2}

\section{Codes On Graphs}
The prime example of codes on graphs are LDPC codes. The development of the field is so interlinked with LDPC codes that it is difficult to distinguish between them.\\
About five years ago, the field was reignited which was dormant from the time of Gallager \cite{a1} who propounded the idea of LDPC codes and iterative techniques. This was largely due to independent discovery of power and efficiency of LDPC codes by several researchers and on the other was the pioneering thesis of Wiberg \cite{a3}. His most important contribution may have been to extend the Tanner graphs to include the state variables as well as as symbol variables.\\
The LDPC decoding algorithms maybe understood as instances of iterative sum-product decoding applied to factor graphs. For cycle-free factor graphs, the sum-product algorithm arises as a natural local message passing algorithm for computing function "summaries", analogous to marginal probabilities. For more on progress on graphs refer to \cite{a17}.

\chapter{A New Decoding Algorithm Utilizing the turbo principle}
\begin{figure}
\begin{center}
  \scalebox{0.5}{\includegraphics{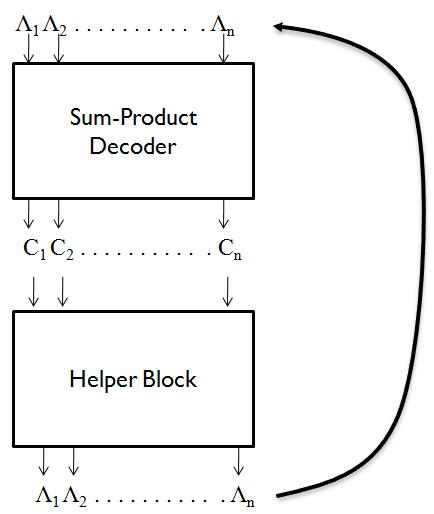}}

  \caption{Idea of the proposed decoding algorithm}\label{fig:fig4}
  \end{center}
\end{figure}

In this chapter, we develop the ideas of a new decoding method utilizing the state estimation data and even how to estimate the necessary parameters for the markov model developed for the state of the channel. Some of the ideas are based on the works of \cite{a4} and \cite{a6} and can be referred.
The proposed decoding method has two blocks. One is the conventional sum-product algorithm part and the other is a helper part utilizing state data.
The helper part evaluates the likelihood ratio conditional to the set of received symbols and the state information.The two parts work in cooperative manner
in an iterative decoding process. The helper part outputs are used as the input to
the sum-product algorithm. After performing the sum-product computation part, we obtain the first tentative word $\hat{\textbf{c}}$.
From $\hat{\textbf{c}}$, the helper part updates
the Log Likelihood Ratio(LLR). Then, the next round of the sum-product algorithm is started. In this new round, the updated LLRs are used
as the input. The two processes (i.e., the helper part and
the sum-product algorithm) are repeated alternately until the correct decision is reached or the max no of iterations are reached.
Figure \ref{fig:fig4} illustrates the idea of the
proposed decoding algorithm.
More about the sequence of operations is discussed in section 3.3 which develops these ideas once the algorithm has been described.
\section{Markov Noise Channel}
\begin{figure}
\begin{center}
  \scalebox{0.7}{\includegraphics{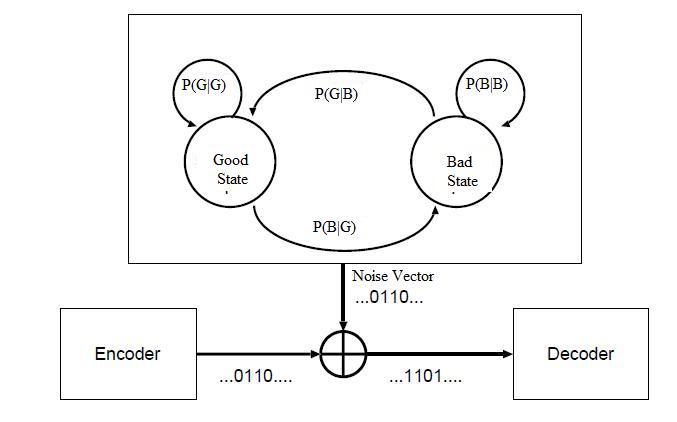}}

  \caption{Two-state hidden Markov noise channel}\label{fig:fig5}
  \end{center}
\end{figure}

A markov channel is defined by a set of states and transition probabilities between those states. Each state will have characteristic noise associated. In this work, we will assume bit flipping noise. Each bit transmitted will have a state associated with it and the channel state for the next transmission is determined by the state transition probability. Each state therefore has a typical run length in state sequence.
Thus here we only treat the cases where the noise is binary in nature. The noise channel is additive noise channel. Hence the received vector is given by $y = x \oplus z$, where $x \in C$ and $z \in \{0,1\}^{N}$; the statistics of the noise vector $z$ is described by a Markov Model (MM). Thus the output of MM is a bit flipping noise. The two state markov noise channel is illustrated in figure \ref{fig:fig5}.
The state space of the MM is denoted by $S \triangleq \{0,1,...,s_{max}\}$. The
state transition from state $j$ to state $i$ occurs with the probability $p(i|j)$ where $i,j \in (G,B)$. We are considering GEC model which is a two state MM with a good state G and a bad state B. Refer to introduction for more details.
At the transition form state $j$ to state $i$, the MM emits symbol 1 with
the probability $q_{i \leftarrow j}$ and symbols 0 with the probability 1 - $q_{i \leftarrow j}$. These are the parameters which characterize a MM. Later we will descibe a method to estimate these parameters as well (see section 3.4).

\section{Turbo Algorithm for computing helper Log-Likelihood Ratios}
The logrithm of the likelihood ratio $\Lambda(d_{k})$ associated with each of the bit $d_{k}$ is given by
\begin{equation}
\Lambda(d_{k}) = Log \frac{Pr(d_{k}=1|observation)}{Pr(d_{k}=0|observation)}
\end{equation}
where $Pr(d_{k}=i|observation)$,$i \in \{0,1\}$ is the \emph{aposteriori} probabilities(APP) associated with the bit $d_{k}$. Assuming that the bits $d_{k}$ at the input takes value of 0 and 1 with equal frequency and $O_{1}^{N}$ is the received vector(or $\textbf{O}$), then the APP of the data bit sent $d_{k}$ can be computed by the joint probability function $\zeta(i,m)$ as
\begin{equation}
\zeta(i,m) := Pr(d_{k}=i,S_{k}=m|\textbf{O})
\end{equation}
which is the probability of the transmitted bit being $i$ and the markov channel state being $m$ given the received vector. The APP will be subsequently given by the following simple summation rule:
\begin{equation}
Pr(d_{k}=i|\textbf{O}) = \sum_{m} \zeta(i,m)
\end{equation}
The summation of $\zeta$ over all the possible channel states (two in our case as we have assumed GEC model) gives the required probability.
The helper LLR , in this case, is given by the ratio
\begin{equation}
H\Lambda(d_{k}) = Log \frac{\sum_{m} \zeta(i=1,m)}{\sum_{m} \zeta(i=0,m)}
\end{equation}
Let us introduce some new variables for the calculation of $\zeta(i,m)$. Let us define the following:
\begin{equation}
\alpha_{k}(i,m) := Pr(d_{k}=i,S_{k}=m|O_{1}^{k}) = \frac{Pr(d_{k}=i,S_{k}=m,O_{1}^{k})}{Pr(O_{1}^{k})}
\end{equation}
Here, the observations are taken only upto the instant $k$. Let us go further and introduce two new variables as
\begin{equation}
\beta_{k}(m) := \frac{Pr(O_{k+1}^{N}|S_{k}=m)}{Pr(O_{k+1}^{N}|O_{1}^{k})}
\end{equation}
\begin{equation}
\gamma_{i}(O_{k},m_{1},m_{2}) := Pr(d_{k}=i,O_{k},S_{k}=m_{2}|S_{k-1}=m_{1})
\end{equation}
where $O_{k}$ is the observation at the instant $k$.
Now we can write the joint probability $\zeta(i,m)$ in terms of $\alpha_{k}(i,m)$ and $\beta_{k}(m)$ as follows:

\begin{eqnarray}
  \zeta(i,m) &=&\frac{Pr(d_{k}=i,S_{k}=m,O_{1}^{k},O_{k+1}^{N})}{Pr(O_{1}^{k},O_{k+1}^{N})}\nonumber\\
   &=&\frac{Pr(d_{k}=i,S_{k}=m,O_{1}^{k})}{Pr(O_{1}^{k})} . \frac{Pr(O_{k+1}^{N}|d_{k}=i,S_{k}=m,O_{1}^{k})}{Pr(O_{k+1}^{N}|O_{1}^{k})}
\end{eqnarray}

The observations after the time $k$ is not influenced either by $d_{k}$ or by observations upto time $k$ but only by $S_{k}$, thus $\zeta(i,m)$ may be written as
\begin{equation}
\zeta(i,m) = \alpha_{k}(i,m) . \beta_{k}(m)
\end{equation}
The probabilities $\alpha_{k}(i,m)$ and $\beta_{k}(m)$ can be recursively calculated using the following recursions which are proved in the appendix of \cite{a6}. It uses the $\gamma_{i}(O_{k},m_{1},m_{2})$ probability in the following way:
\begin{equation}\label{e:eq10}
\alpha_{k}(i,m) = \frac{\sum_{m_{1}} \sum_{j} \gamma_{i}(O_{k},m_{1},m_{2}) \alpha_{k-1}(j,m_{1})}{\sum_{m_{1}}\sum_{m_{2}}\sum_{i}\sum_{j} \gamma_{i}(O_{k},m_{1},m_{2}) \alpha_{k-1}(j,m_{1})}
\end{equation}
and
\begin{equation}\label{e:eq11}
\beta_{k}(m) = \frac{\sum_{m_{1}} \sum_{j} \gamma_{i}(O_{k+1},m_{2},m_{1}) \beta_{k+1}(m_{1})}{\sum_{m_{1}}\sum_{m_{2}}\sum_{i}\sum_{j} \gamma_{i}(O_{k+1},m_{1},m_{2}) \alpha_{k}(j,m_{1})}
\end{equation}
The probability $\gamma_{i}(O_{k},m_{1},m_{2})$ can be computed using the parameters of the noise channel model that we have proposed. Using the definition of
$\gamma_{i}(O_{k},m_{1},m_{2})$ that we have given earlier, we can write


\begin{eqnarray}\label{e:eq51}
  \gamma_{i}(O_{k},m_{1},m_{2}) &=&Pr(O_{k}|d_{k}=i,S_{k}=m_{2},S_{k-1}=m_{1}).\nonumber\\
   &&Pr(d_{k}=i|S_{k}=m_{2},S_{k-1}=m_{1}). \nonumber\\
   &&Pr(S_{k}=m_{2}|S_{k-1}=m_{1})
\end{eqnarray}

Here, $Pr(O_{k}|d_{k}=i,S_{k}=m_{2},S_{k-1}=m_{1})$ corresponds to the error probability if $y \neq i$ in the given state of the channel and probability of correct reception if $y = i,i \in (0,1)$. In terms of the channel parameters, it corresponds to $q_{i \leftarrow j}$ or 1 - $q_{i \leftarrow j}$ depending upon whether the observed data at the $k^{th}$ instant is complement of $i$ or $i$ itself. The next term $Pr(d_{k}=i|S_{k}=m_{2},S_{k-1}=m_{1})$ is 0.5 as the probability of transmission of any $i^{th}$ symbol is equal uniform for any given state. Since there are only two symbols, it comes out to be 1/2. The \emph{apriori} probabilities does not depend upon the state. The third term $Pr(S_{k}=m_{2}|S_{k-1}=m_{1})$ is the state transition probability and is the probability of channel to transition to state $m_{2}$ given that the current state is $m_{1}$. In terms of the state parameters, this can be written as
$p(m_{2}|m_{1})$ where $m_{1}$ and $m_{2} \in (G,B)$. Thus we can write the above equation in terms of channel parameters or parameters of MM as:
%

\begin{eqnarray}
  \gamma_{i}(O_{k},m_{1},m_{2}) &=&0.5.(q_{i \leftarrow j}).p(m_{2}|m_{1}) \hspace{.15cm}if \hspace{.15cm}O_{k} \neq i \nonumber\\
   &=&0.5.(1-q_{i \leftarrow j}).p(m_{2}|m_{1}) \hspace{.15cm}if \hspace{.15cm}O_{k} = i
\end{eqnarray}
Here $(i,j,m_{1},m_{2}) \in (G,B)$. Thus the various steps in the computations of APP for all $d_{k}$ are given below:
\begin{itemize}
  \item The probabilities $\alpha_{0}(i,m)$ and $\beta_{k}(m)$ are initialized in following way:
  \begin{equation}
  \alpha_{0}(i,0) = 1 , \alpha_{0}(i,m) = 0 \hspace{.15cm}for \hspace{.15cm}all \hspace{.15cm}m \neq 0
  \end{equation}
  This comes from the fact that we always start with the good state, otherwise initialize the one that you are starting with. In the similar manner,
  \begin{equation}
  \beta_{k}(0) = 1 , \beta_{k}(m) = 0 \hspace{.15cm}for \hspace{.15cm}all \hspace{.15cm}m \neq 0
  \end{equation}
  \item For each of the observations $O_{k}$, we compute $\alpha_{k}(i,m)$ and $\gamma_{i}(O_{k},m_{1},m_{2})$ in accordance with equations \ref{e:eq10} and \ref{e:eq51} respectively.
  \item After the full sequence has arrived, thus we have the full vector \textbf{O}, then $\beta_{k}(m)$ can be calculated in accordance with the equation \ref{e:eq11}.
  \item Now both the $\alpha_{k}(i,m)$ and $\beta_{k}(m)$ can be multiplied together to obtain $\zeta(i,m)$ for each bit. Then APP for each bit can then be obtained.
\end{itemize}

\section{Message Passing Schedules}
A brute force method to minimizing the word or symbol error probability would certainly try to minimize the probability $Pr(\textbf{x}|\textbf{y})$ for each of the codeword $\textbf{x}$. The complexity of the brute force method is proportional to the no. of codewords $\sim 2^{k}$. The algorithm derived here tries to obtain the APP of the information and channel digits leading to a soft decoder instead of a hard decoder.
\subsection{Standard Message passing schedules and Local neighborhoods}
\begin{figure}
\begin{center}
  \scalebox{0.7}{\includegraphics{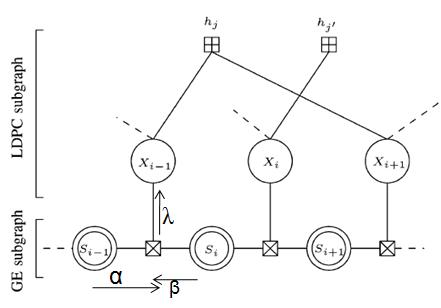}}

  \caption{Factor graph of a combined LDPC code and markov modeled noise channel}\label{fig:fig50}
  \end{center}
\end{figure}

Estimation decoding in GEC channel requires not only the sum-product calculations, but also a message passing schedule, which defines the order in which the calculations occur. What we have used here is a standard message passing schedule. Here, first perform the sum-product calculations at the symbol nodes and then at the check nodes, then the tentative word is passed to helper block which calculates the APP ratios for each bit. This acts as an extrinsic information at the symbol nodes and the sum-product calculation at the next iteration takes care of this extrinsic information.\\
The decoding strategy considered here is equivalent to working on a factor graph with messages as shown in figure \ref{fig:fig50}.

To analyze the messages passed under this decoding scheme, we will form a subgraph of the overall graph of the factor graph containing all nodes and edges that participate in the calculation of a particular message along edge $e$ at iteration no. $l$. This subgraph is called Local neighborhood of the edge $e$ and written as $\aleph_{e}(l)$.\\
There can be different message passing schedules which gives rise to different nodes and edges in the calculation and hence gives rise to different local neighborhood. The operation that makes a complete iteration in the standard schedule can be divided into:
\begin{itemize}
  \item LDPC sub-iteration, which represents message passing operations at the parity check nodes and at the symbol variable nodes.
  \item Helper block sub-iteration, representing the message passing operations at the channel state and factor nodes.
\end{itemize}

\section{Channel Parameter estimation by using Baum-Welch Algorithm}
The knowledge of channel parameters of MM is necessary to use this
proposed decoding algorithm. However, in a practical situation, the channel
parameters may not be available a priori. Thus an online estimation method has to be
constructed based on an algorithm for estimating the channel parameters. This can be done using an expectation maximization technique called the \emph{Baum-Welch} algorithm.\\
This is a method to adjust the model parameters to maximize the probability of observation sequence given the model. We can choose $\lambda$ , as the model such that $p(\textbf{O}|\lambda)$ is locally maximized with Baum-Welch algorithm. Let us start defining the procedure for estimation of MM parameters. For this, we need to define $\upsilon_{t}(m_{1},m_{2})$, the joint probability of being in state $m_{1}$ at time $k$ and $m_{2}$ at time instant $k+1$, given the model and the observation sequence. Thus, it can be written as:
\begin{equation}
\upsilon_{k}(m_{1},m_{2}) := Pr(S_{k}=m_{1},S_{k+1}=m_{2}|\textbf{O},\lambda)
\end{equation}
Now, we define forward and backward variables as follows: The forward variable $\alpha_{t}(m)$ is the joint probability of partial observation sequence $O_{1}^{k}$ and state $S_{k}$ at time $k$ given the model $\lambda$.
\begin{equation}
\alpha_{k}(m) := Pr(O_{1}^{k},S_{k}=m|\lambda)
\end{equation}
In the similar manner, we may define the backward variable $\beta_{k}(m)$ as following:
\begin{equation}
\beta_{k}(m) := Pr(O_{k+1}^{N}|S_{k}=m,\lambda)
\end{equation}
It is the probability of partial observation sequence from $k+1$ till the end given state $S_{k}$ at time $k$ and model $\lambda$. Thus, from the definitions of the forward and the backward variables, we can write $\upsilon_{k}(m_{1},m_{2})$ in terms of the parameters of the model as below \cite{a5}. Here, $b_{m}(O_{k})$ is the probability of observing a particular symbol $\in (0,1)$ in state $m$.
\begin{equation}
\upsilon_{k}(m_{1},m_{2}) := \frac{\alpha_{k}(m).p(m_{1}|m_{2}).b_{m_{2}}(O_{k+1}).\beta_{k+1}(m_{2})}{Pr(\textbf{O}|\lambda)}
\end{equation}
where $b_{m}(O_{k}$ can be $q_{i \leftarrow j}$ or (1-$q_{i \leftarrow j}$) depending upon the received symbol at time $k$.
Let us define $\eta_{k}(m)$ as the probability of being in state $m$ at time $k$ , given the observation sequence and the model $\lambda$, then we can relate $\eta_{k}(m)$ to $\upsilon_{k}(m_{1},m_{2})$ as
\begin{equation}
\eta_{k}(m_{1}) = \sum_{m_{2}} \upsilon_{k}(m_{1},m_{2})
\end{equation}
If we sum $\eta_{k}(m_{1})$ over the time index $k$ we get the quantity which can be interpreted as the expected no. of times the state $m_{1}$ is visited. Similarly, the summation over $k$ of $\upsilon_{k}(m_{1},m_{2})$ can be interpreted as the expected no. of transitions from state $m_{1}$ to $m_{2}$. Thus,
\begin{equation}
\sum_{k} \eta_{k}(m_{1}) = expected \hspace{.15cm}no. \hspace{.15cm}of \hspace{.15cm}transitions \hspace{.15cm}from \hspace{.15cm}state \hspace{.15cm}m_{1}
\end{equation}
\begin{equation}
\sum_{k} \upsilon_{k}(m_{1},m_{2}) = expected \hspace{.15cm}no. \hspace{.15cm}of \hspace{.15cm}transitions \hspace{.15cm}from \hspace{.15cm}state \hspace{.15cm}m_{1} \hspace{.15cm}to \hspace{.15cm}state \hspace{.15cm}m_{2}.
\end{equation}
Using these above formulas, we can give a method for estimation of parameters of MM. These formulas are stated below:
\begin{equation}
p(i|j) = \frac{expected \hspace{.15cm}no. \hspace{.15cm}of \hspace{.15cm}transitions \hspace{.15cm}from \hspace{.15cm}state \hspace{.15cm}m_{1} \hspace{.15cm}to \hspace{.15cm}state \hspace{.15cm}m_{2}}{expected \hspace{.15cm}no. \hspace{.15cm}of \hspace{.15cm}transitions \hspace{.15cm}from \hspace{.15cm}state \hspace{.15cm}m_{1}}.
\end{equation}
Thus,
\begin{equation}
p(i|j) = \frac{\sum_{k} \upsilon_{k}(m_{1},m_{2})}{\sum_{k} \eta_{k}(m_{1})}
\end{equation}
Here, $m_{1},m_{2} \in (G,B)$ and $i,j \in (0,1)$Similarly,
\begin{equation}
b_{m}(O_{k}) = \frac{expected \hspace{.15cm}no. \hspace{.15cm}of \hspace{.15cm}times \hspace{.15cm}in \hspace{.15cm}state \hspace{.15cm}m \hspace{.15cm}and \hspace{.15cm}observing \hspace{.15cm}symbol \hspace{.15cm}O_{k}}{expected \hspace{.15cm}no. \hspace{.15cm}of \hspace{.15cm}times \hspace{.15cm}in \hspace{.15cm}state \hspace{.15cm}m}
\end{equation}
Hence,
\begin{equation}
b_{m}(O_{k}) = \frac{\sum_{k \hspace{.1cm}s.t \hspace{.1cm}O_{k}} \eta_{k}(m)}{\sum_{k} \eta_{k}(m)}
\end{equation}
If we define the current model as $\lambda$ and use the above set of equations and define the re-estimated model as $\overline{\lambda}$, then it has been proven by Baum and his colleagues \cite{a16} that
\begin{itemize}
  \item Either the initial model $\lambda$ defines the critical point of the likelihood function, or,
  \item model $\overline{\lambda}$ is more likely than the model $\lambda$ in the sense that $p(\textbf{O}|\overline{\lambda}) > p(\textbf{O}|\lambda)$, ie, we have found a new model $\overline{\lambda}$ from which the observation sequence is likely to have been produced.
\end{itemize}
Based on the above procedure, if we iteratively use $\overline{\lambda}$ in place of $\lambda$, then we can improve the probability of $\textbf{O}$ being observed from the model until some limiting point is reached. The final result is called a maximum likelihood estimate of MM.

\section{Designing good LDPC codes}
An LDPC code design problem can be posed as follows:\\
Given a particular Comm channel\\
\hspace{.3cm}\textbf{Max.} Code Rate\\
\hspace{.3cm}\textbf{Subject to} $P_{err} < \epsilon$,\\
where $P_{err}$ is the probability of symbol error and $\epsilon$ is the max. acceptable probability of error \cite{a13}. LDPC codes maybe characterized by degree sequences, which expresses the probability of finding a given no. of ones in either a row or column of a parity check matrix. For a given degree sequence, the symbol wise error performance of the LDPC code for a markov channel in the limit of long block length requires a modified density evolution taking state estimation into consideration, which can be used to verify the probability of error criterion. This task is more elaborated in \cite{a13}.

\chapter{Gallager's Decoding Algorithm Revisited}
\begin{figure}
\begin{center}
  \scalebox{0.7}{\includegraphics{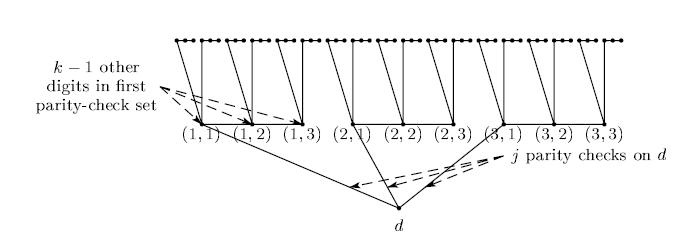}}

  \caption{Parity Check Set tree}\label{fig:fig6}
  \end{center}
\end{figure}

The best decoding strategy is to find the maximum likelihood function by using the minimum distance criterion between the received vector and the transmitted waveform. However, implementing such a receiver would be a big task as the code length increases. LDPC codes achieves good performances as the code length becomes larger and larger. So for such codes, implementing a maximum likelihood receiver is practically not feasible. Gallager has devised many decoding methods in his doctoral thesis \cite{a1}. Here what we are going to describe is the modification of probabilistic decoding method (proposed in his thesis way back in the 60's and not used later) and how the state estimation (or state information) can dramatically improve the decoding performance of this scheme compared to the memoryless case. This scheme makes use of the parity check structure and the digits seeming to be unconnected with the digit being decoded in a very systematic way. It is an iterative technique which tries to correct some errors in the first iteration and based on the corrected digits in the first round, tries to correct some more in the next and so on.\\
Before going further into details , let us understand the basic tree structure required for our decoding purpose. This will also aid us in understanding how an arbitrary digit $d_{k}$
can be corrected even if its parity check sets contain more than one transmission error. Digit $d_{k}$ is represented by the node at the base of the tree and each line rising from this node represents one of the parity check sets containing digit $d_{k}$. The other digits are represented by the nodes on the first tier of the tree. The lines rising from tier 1 to tier 2 of the tree represents the other parity checks containing the digits on tier 1 and the nodes in tier 2 represents the other digits in those parity check sets. Here we are considering decoding of $(n,j,k)$ codes , where $n$ is the length of the code, $j$ denotes the no. of parity check sets of a digit and $k$ denotes the no. of digits contained in each of the parity check set. Keeping these in mind, the tree structure just described for an $(n,3,4)$ code would look like the one shown in figure \ref{fig:fig6}.

If we consider that several of the digits in the first tier are in error, then on the first decoding attempt, the error free digits in the second tier and their parity check equations will allow correction of errors in the first tier. This in turn will allow correction of the digit $d_{k}$ in the second decoding attempt.
\section{Probabilistic Decoding}
\begin{figure}
\begin{center}
  \scalebox{0.7}{\includegraphics{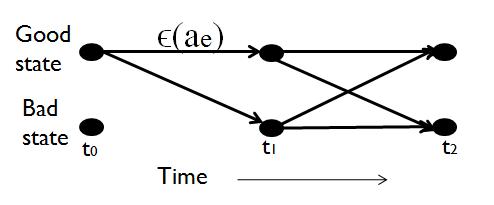}}

  \caption{Trellis Diagram}\label{fig:fig7}
  \end{center}
\end{figure}

Here we derive an iterative procedure will be evolved that on the $l^{th}$ iteration computes the probability that the transmitted digit in the position $d_{k}$ is a 1 conditional on the received symbols out to and including the $l^{th}$ tier. \\
Consider an ensemble of events in which the transmitted digits in the positions of $d_{k}$ and the first tier are independent equiprobable binary digits.
Within this ensemble, we want to find out the probability that the transmitted digit is a 1 conditional on the set of received symbols {\textbf{y}}, the event $S$ that the transmitted digits satisfy the $j$ parity check equations on $d_{k}$ and the start state and end state defined next. Let us have an event that at the given instant the channel transitions from a start state (StState) to an end state(EState) and gives out an error symbol with probability $q_{EState \leftarrow StState}$ where StState and EState $\in (G,B)$. This is in accordance with the definition that we have given earlier for MM noise channel. Here the channel model is
\begin{equation}
\textbf{y} = \textbf{x} \oplus \textbf{n}
\end{equation}
where the noise vector is being generated from the MM noisy channel. We have already estimated the channel parameters (for details refer to section 3.4 in last chapter). This event can also be represented in the form of trellis where there are nodes for the states and on the horizontal axis, we have the time. Hence we have two nodes corresponding to the good state and the bad state of GEC model (refer to introduction chap 1). A branch is said to connect a given start node to an end node if there exists a transition from that start state to end state. Each branch is associated with the error probability as illustrated in the figure \ref{fig:fig7}. Thus we have the transitions starting at nodes $\in (G,B)$ and ending at nodes $\in (G,B)$ as the time progresses.

Hence, given the input, the channel gives an output and then transitions to the other state. This event is synonymous to traversing a branch of the trellis. Let the probability of traversing a branch be given by $\eta$. This probability can be obtained from the parameters of the MM as
\begin{equation}\label{e:eq14}
 \eta = P_{o}(\epsilon|StState) . p(EState|StState)
 \end{equation}
 assuming independence between the giving out of the output and transition occurring thereafter.
  Here, $P_{o}(\epsilon|StState)$ gives the probability of producing an output given the state. It is same as $q_{EState \leftarrow StState}$. $p(EState|StState)$ is the transition probability from Start state to End State.\\
 Using this ensemble, we want to find the APP of the digit as :
 \begin{equation}
 Pr(x_{d}=1|\textbf{y},S,StState,EState)
 \end{equation}
 Let $\eta_{d_{k}}$ be the probability that the transmitted digit in position $d_{k}$ is a 1 conditional to the branch being traversed in the trellis, and let $\eta_{il}$ be the same probability for the $l^{th}$ digit in the $i^{th}$ parity check set of the first tier. Then,
%

\begin{eqnarray}\label{e:eq15}
\frac{Pr(x_{d_{k}}=0|\textbf{y},S,StState,EState)}{Pr(x_{d_{k}}=1|\textbf{y},S,StState,EState)} &=&\frac{1-\eta_{d_{k}}}{\eta_{d_{k}}}.\prod_{i=1}^{j}\frac{(1+\prod_{l=1}^{k-1}(1-2\eta_{il}))}{(1-\prod_{l=1}^{k-1}(1-2\eta_{il}))}\nonumber\\
&=&\frac{1-q_{EState \leftarrow StState}.p(EState|StState)}{q_{EState \leftarrow StState}.p(EState|StState)}.\\
&&\prod_{i=1}^{j}\frac{(1+\prod_{l=1}^{k-1}(1-2q_{EState \leftarrow StState}.p(EState|StState)))}{(1-\prod_{l=1}^{k-1}(1-2q_{EState \leftarrow StState}.p(EState|StState)))}\nonumber
\end{eqnarray}

 Here, $\eta_{d_{k}}$ or $\eta_{il}$ is replaced by $\eta$ as given in equation \ref{e:eq14}. To prove the above, we need the following lemma:\\
 \textbf{\emph{Lemma}}: If we consider a set of $m$ independent digits with the probability of occurrence of a 1 at location $l$ being given by $P_{l}$, then the probability of occurrence of an even no. of ones is given by:
 \begin{equation}
 \frac{1+\prod_{l=1}^{m}(1-2P_{l})}{2}
 \end{equation}
 For proof of this lemma, refer to \cite{a1}. Let us consider the following statements about conditional probabilities:
%
%
%

 \begin{eqnarray}
  \frac{Pr(x_{d_{k}}=0|\textbf{y},S,StState,EState)}{Pr(x_{d_{k}}=1|\textbf{y},S,StState,EState)}
  &=&\frac{Pr(x_{d_{k}}=0,\textbf{y},S,StState,EState)}{Pr(x_{d_{k}}=1,\textbf{y},S,StState,EState)}\nonumber\\
   &&\hspace{-6.5cm} \prod_{i}\frac{Pr(S_{i}|x_{d_{k}}=0,\textbf{y},StState,EState)}{Pr(S_{i}|x_{d_{k}}=1,\textbf{y},StState,EState)}.\frac{Pr(x_{d_{k}}=0,\textbf{y},StState,EState)}{Pr(x_{d_{k}}=1,\textbf{y},StState,EState)}\nonumber\\
   &&\hspace{-6.5cm} \prod_{i}\frac{Pr(S_{i}|x_{d_{k}}=0,\textbf{y},StState,EState)}{Pr(S_{i}|x_{d_{k}}=1,\textbf{y},StState,EState)}.\frac{Pr(x_{d_{k}}=0|\textbf{y},StState,EState)}{Pr(x_{d_{k}}=1|\textbf{y},StState,EState)}
\end{eqnarray}

 Thus,
 \begin{equation}\label{e:eq55}
 \frac{Pr(x_{d_{k}}=0|\textbf{y},S,StState,EState)}{Pr(x_{d_{k}}=1|\textbf{y},S,StState,EState)}
 =\frac{1-\eta_{d_{k}}}{\eta_{d_{k}}}.\prod_{i}\frac{Pr(S_{i}|x_{d_{k}}=0,\textbf{y},StState,EState)}{Pr(S_{i}|x_{d_{k}}=1,\textbf{y},StState,EState)}
 \end{equation}
 If it is given that $x_{d_{k}}=0$, then the parity check on $d_{k}$ is satisfied if the other $k-1$ digits contain an even no. of ones. Using the Lemma, this probability is equal to
 \begin{equation}
 Pr(S|x_{d_{k}}=0,\textbf{y},StState,EState) = \prod_{i=1}^{j}\frac{1+\prod_{l=1}^{k-1}(1-2\eta_{il})}{2}
 \end{equation}
 Similarly,
 \begin{equation}
 Pr(S|x_{d_{k}}=1,\textbf{y},StState,EState) = \prod_{i=1}^{j}\frac{1-\prod_{l=1}^{k-1}(1-2\eta_{il})}{2}
 \end{equation}
 Substituting the above equations into equation \ref{e:eq55} , we get the required result. This gives us the likelihood ratio for the digit $d_{k}$. The decoding procedure for the entire code can now be stated as follows:
 \section{Decoding Procedure}
 This remains the same as prescribed in \cite{a1}. It is restated here for clarity.
 For each digit and each combination of $j-1$ parity check sets containing that digit, we use equation \ref{e:eq15} to calculate the probability of a transmitted 1 conditional to the branch traversed in the trellis in the $j-1$ parity check sets. Thus, there are $j$ different probabilities associated with each digit. Next these probabilities are used in equation \ref{e:eq15} to calculate a next set of probabilities. The probability to be associated with one digit in the calculation of another digit $d$ is the probability found in the first iteration, omitting the parity check set containing $d$. If the decoding is successful, then the decoding associated with each digit tends to 0 or 1 as the no. of iterations is increased. The calculations are valid as long as the independence assumption holds otherwise the tree would close upon itself.\\
 The convergence issue being highlighted here is that of probability of decoding a digit $d_{k} \rightarrow 0,1$ as no. of iteration increases has not been addressed here and can be a topic for future study. This has been simply stated in \cite{a1} and this has not been proved there as well.
 \section{Convergence of decoding error probability of a symbol towards zero for given channel parameters}
 Density evolution (DE) analysis provides performance thresholds for LDPC codes, establishing a region of channel parameters over which iterative decoding is successful in the limit of long block length or as the no. of iterations is increased. The DE algorithm establishes bound on the ultimate performance of LDPC codes allowing a calculation of a threshold in memoryless channels. This is a fairly complicated algorithm utilizing many approximations (such as gaussian approximation for densities of messages at the nodes). This has been an area of active research to speed up the process of DE calculations as DE is utilized in various other tasks like designing good LDPC codes etc. Here we are proposing an approximation of DE algorithm to check whether given channel parameters can lead to successful decoding or not. This is fairly fast and simple strategy derived from the work of \cite{a1}. The DE algorithm implements a decision function, taking channel parameters as inputs, and determining whether or not the decoder achieves a very small probability of error for these parameter values. Our work focusses on a approximate DE based analysis of LDPC decoding over the GE channel. \\
 The channel model is the same as introduced before in last chapter (refer to section 3.1). It is a binary input, binary output channel in which the channel output $\textbf{y} \in \{0,1\}^{n}$ in response to the channel input vector $X \in \{0,1\}^{n}$ is given by
 \begin{equation}
 \textbf{y} = \textbf{x} \oplus \textbf{z}
 \end{equation}
 where $z \in \{0,1\}^{n}$ is a noise sequence and $\oplus$ denotes componentwise modulo 2 addition.The noise sequence arises from a two state MM channel. The transition probabilities are given by $p(i|j)$ which tells us of the probability of transition from state $j$ to state $i$ where $i$ and $j \in \{G,B\}$. The inversion probabilities in these states is given by $q_{EState \leftarrow StState}$ where StState and EState $\in \{G,B\}$. The average inversion probability is given by:
 \begin{equation}\label{e:eq16}
 \overline{\eta} := Pr(z_{i}=1) = \frac{p(B|G).q_{B \leftarrow G} + p(G|B).q_{G \leftarrow B}}{p(B|G) + p(G|B)}
 \end{equation}
We assume an $(n,j,k)$ code with $j=3$ parity check sets constraining each of the digit. Let the tree consist of $l$ independent tiers with the uppermost tier be labeled as tier $0$ and the bottom most one is tier $m$. Here, let the decoding procedure be if both the parity check fails for a digit, then change the digit (because of the strong indication of error of the digit as both parity checks have failed). Then using these changed digits in the first tier, we calculate the parity checks in the second tier and repeat the same procedure. We now try to determine the probability of decoding error for the digit in tier $l$. If the digit is received in error (an event of probability $\overline{\eta}$), then a parity check set will be satisfied if there is an even no. of errors in the remaining $k-1$ digits. From the Lemma given earlier, this probability is equivalent to:
\begin{equation}
\frac{1 + (1-2\overline{\eta})^{k-1}}{2}
\end{equation}
Following similar footsteps as in \cite{a1}, we arrive at the following conclusion: If $\overline{\eta}_{i}$ is the probability of error after processing of a digit in the $i^{th}$ tier, then
\begin{equation}\label{e:eq17}
\overline{\eta}_{i+1} = \overline{\eta} - \overline{\eta}(\frac{1+(1-2\overline{\eta}_{i})^{k-1}}{2})^{2} + (1-\overline{\eta})(\frac{1-(1-2\overline{\eta}_{i})^{k-1}}{2})^{2}
\end{equation}
Here $\overline{\eta}$ is defined in equation \ref{e:eq16}. Using the approximate DE algorithm as given in equation \ref{e:eq17}, we try to locate points (which are two tuples ($q_{B \leftarrow G},q_{G \leftarrow B}$)) where the probability of error for a symbol $\rightarrow 0$ as the no. of iterations are increased. Thus, we get a region within which the inversion (or error) probability converges to zero which we refer to as the decoding region. Within this region, choosing the channel parameters leads to successful decoding. This is a useful concept which conveys a priori whether our choice of channel parameters would lead to successful decoding or not. The simulation results are shown in the chapter 6.

\chapter{An Error Predicting Scheme utilizing the channel memory }
This scheme operates on the received channel symbols $\textbf{y}$ and previously decoded data $\hat{\textbf{x}}$ to estimate the probability that the next channel symbol is in error, conditioned on the previous channel errors. The noise MM is defined in earlier chapters. The error process has memory in the sense that it depends on the underlying state process. When conditioned on the state process, the error process is memoryless,ie,
\begin{equation}
Pr(z_{l}|\textbf{s}_{l}) = \prod_{i=1}^{l} Pr(z_{i}|s_{i})
\end{equation}
The state process is a stationary first order markov process:
\begin{equation}
Pr(s_{l}|\textbf{s}_{l-1}) = Pr(s_{l}|s_{l-1})
\end{equation}
\textbf{Preposition:} The following recursions hold:
\begin{equation}
Pr(z_{l+1}=1|\textbf{z}_{l},s_{0}) = \xi(z_{l},Pr(z_{l}=1|\textbf{z}_{l-1},s_{0}))
\end{equation}
and
\begin{equation}
Pr(z_{l+1}=1|\textbf{z}_{l}) = \xi(z_{l},Pr(z_{l}=1|\textbf{z}_{l-1}))
\end{equation}
where the function $\xi$ is defined by:
\begin{equation}
\xi(0,q) \triangleq q_{G \leftarrow B} + p(B|G)(q_{B \leftarrow G} - q_{G \leftarrow B}) + \mu(q - q_{G \leftarrow B})\frac{1-q_{B \leftarrow G}}{1-q}
\end{equation}
and
\begin{equation}
\xi(1,q) \triangleq q_{G \leftarrow B} + p(B|G)(q_{B \leftarrow G} - q_{G \leftarrow B}) + \mu(q - q_{G \leftarrow B})\frac{q_{B \leftarrow G}}{q}
\end{equation}
Here, $\mu$ is defined to be the channel memory
\begin{equation}
\mu = 1-p(G|B)-p(B|G)
\end{equation}
For proof of the above preposition, refer to appendix of \cite{a8}. \\
The error prediction block helps the sum-product algorithm in a similar manner as the helper block of figure 3.1. The error prediction calculates the error probability keeping in mind the channel parameters. These error prediction forms the log-likelihood ratio which aids the sum-product decoder and is simpler to implement than the forward-backward recursions of chap. 2. The log-likelihood ratio can be computed as follows:
\begin{equation}
\lambda = log\frac{Pr(z_{l}=1)}{1-Pr(z_{l}=1)} \hspace{.15cm} for \hspace{.15cm} y \hspace{.15cm} = \hspace{.15cm} 0
\end{equation}
and
\begin{equation}
\lambda = log\frac{1-Pr(z_{l}=1)}{Pr(z_{l}=1)} \hspace{.15cm} for \hspace{.15cm} y \hspace{.15cm} = \hspace{.15cm} 1
\end{equation}

The decisions are based on these likelihood ratios. The bit wise decisions are computed in this way.

\chapter{Simulation Results}
\section{Creating a Parity Check Matrix}
The set of valid codewords for a linear code can be specified by giving a parity check matrix, $H$, with $M$ rows and $N$ columns. The valid codewords are the vectors, $\textbf{x}$, of length $N$, for which $Hx=0$, where all arithmetic is done modulo-2. Each row of $H$ represents a parity check on a subset of the bits in $x$; all these parity checks must be satisfied for $x$ to be a codeword. LDPC codes can be constructed by various methods, which generally involve some random selection of where to put 1s in a parity check matrix. Any such method for constructing LDPC codes will have the property that it produces parity check matrices in which the number of 1s in a column is approximately the same (perhaps on average) for any size parity check matrix. The creation of the parity check matrix involves some precautions which are listed below:
\begin{itemize}
  \item Add 1s to the parity check matrix in order to avoid rows that have no 1s in them, and hence are redundant, or which have only one 1 in them, in which case the corresponding codeword bits will always be zero. The places within such a row to add these 1s are selected randomly.
  \item If the preliminary parity check matrix constructed in step (1) had an even number of 1s in each column, add further 1s to avoid the problem that this will cause the rows to add to zero, and hence at least one check will be redundant. Up to two 1s are added (since it is also undesirable for the sum of the rows to have only one 1 in it), at positions selected randomly from the entire matrix.
  \item We also try to eliminate situations where a pair of columns both have 1s in a particular pair of rows, which correspond to cycles of length four in the factor graph of the parity check matrix. When such a situation is detected, one of the 1s involved is moved randomly within its column.
\end{itemize}

\section{Encoding Message Blocks}
Given a codeword $\textbf{u}$ and an $M \times N$ parity check matrix $H$, we have  $u.H^{t}=0$.  Assume that the message bits $\textbf{s}$ occupy the end of the codeword and the check bits $\textbf{c}$ occupy beginning of the codeword ie, $\textbf{u}=[\textbf{c}|\textbf{s}]$.\\
Let $H=[A|B]$ where $A$ is $M \times M$ matrix and $B$ is $M \times (M-N)$ matrix. The first part of $H$ is an identity matrix . Thus we have
\begin{equation}
Ac+Bs = 0 \hspace{.15cm}or \hspace{.15cm}c = A^{-1}Bs.
\end{equation}
This can be used to compute check bits as long as $A$ is non-singular.\\
After the messages have been encoded, a bit flipping markov model for the noise is simulated with two states \{G,B\} with the error probability in the good state being less than the error probability in the bad state. In our simulations, we have fixed the error probability of the bad state to be equal to $0.5$. The error probability in the good state is varied to vary  the average inversion probability. We have also fixed the error probabilities in the two states and varied the transition probabilities so as the average inversion probability gets altered. The output of the channel is the received vector $\textbf{y}$ and is fed as input to the decoding algorithms. Comparisons are shown in the figures below between three algorithms:
\begin{itemize}
  \item Sum Product algorithm.
  \item Sum-product decoder along with the helper block implemented.
  \item Gallager's probabilistic decoding with side information.
\end{itemize}

\section{Description of figures}
\begin{figure}
\begin{center}
  \scalebox{0.3}{\includegraphics{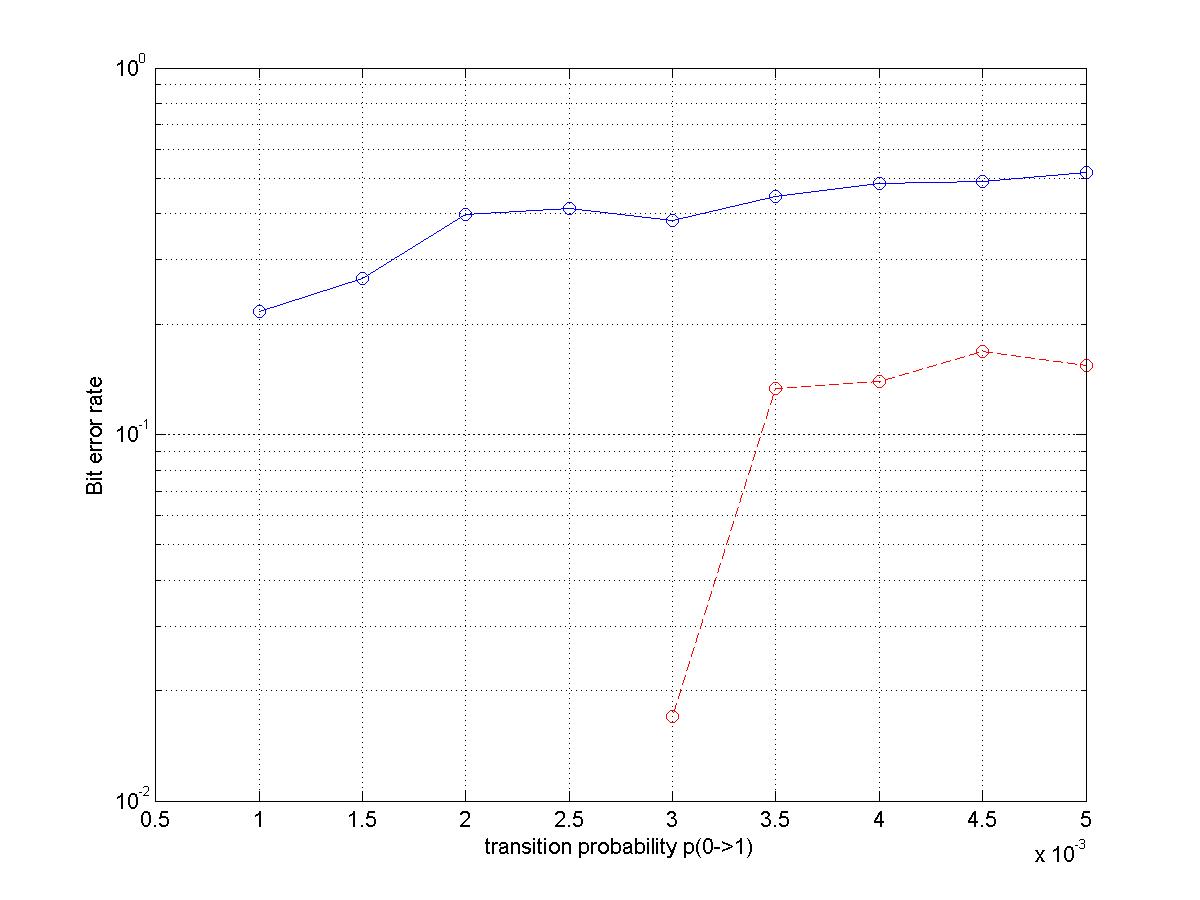}}

  \caption{Comparison between sum-product decoder and sum-product decoder with helper block. N=2000,R=1/2,50 iterations }\label{fig:fig10}
  \end{center}
\end{figure}

\begin{figure}
\begin{center}
  \scalebox{0.3}{\includegraphics{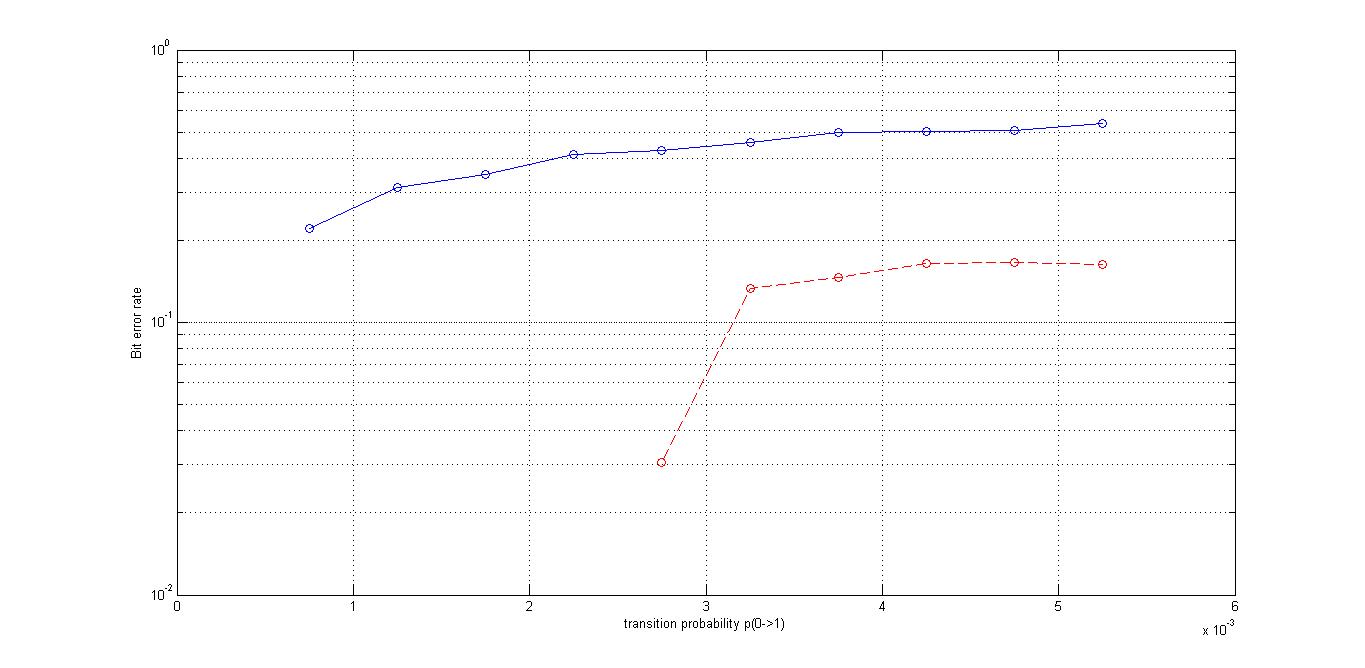}}

  \caption{Comparison between sum-product decoder and sum-product decoder with helper block. N=4000,R=1/2,50 iterations}\label{fig:fig12}
  \end{center}
\end{figure}

\begin{figure}
\begin{center}
  \scalebox{1.0}{\includegraphics{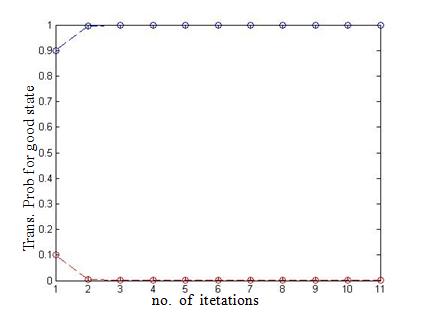}}

  \caption{Convergence of transition probabilities using Baum-Welch algorithm}\label{fig:fig13}
  \end{center}
\end{figure}

\begin{figure}
\begin{center}
  \scalebox{1.0}{\includegraphics{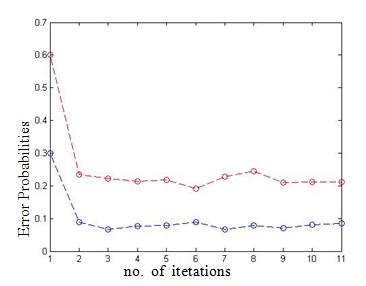}}

  \caption{Convergence of error probabilities using Baum-Welch algorithm}\label{fig:fig14}
  \end{center}
\end{figure}

\begin{figure}
\begin{center}
  \scalebox{1.0}{\includegraphics{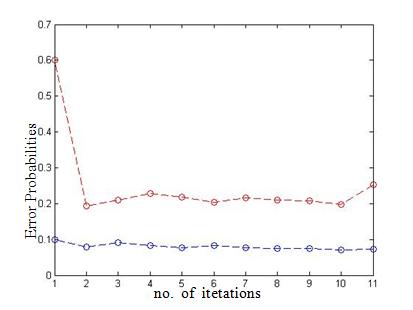}}

  \caption{Convergence of error probabilities in the two states using Baum-Welch algorithm}\label{fig:fig15}
  \end{center}
\end{figure}

\begin{figure}
\begin{center}
  \scalebox{1.0}{\includegraphics{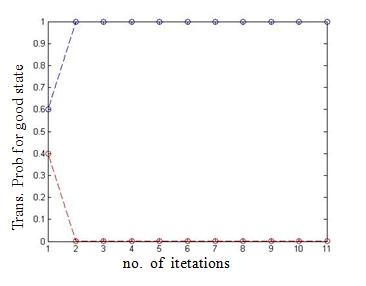}}

  \caption{Convergence of transition probabilities in the two states using Baum-Welch algorithm}\label{fig:fig16}
  \end{center}
\end{figure}

\begin{figure}
\begin{center}
  \scalebox{0.3}{\includegraphics{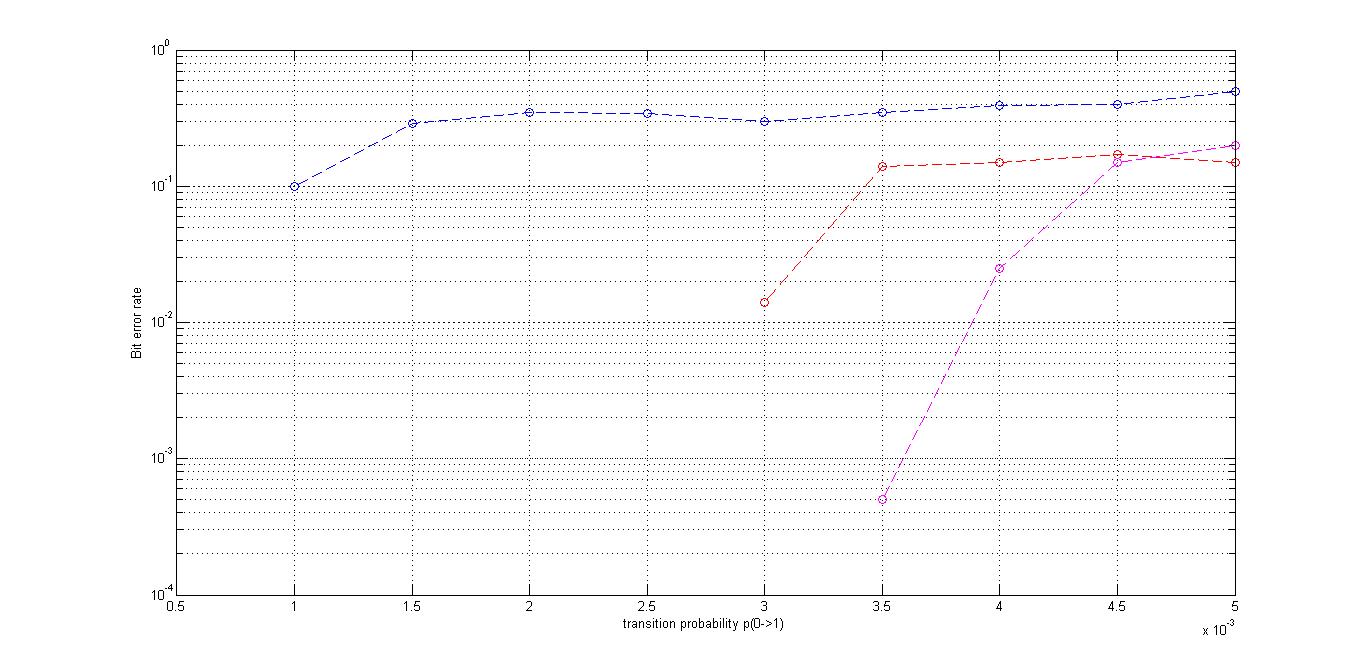}}

  \caption{Comparison of bit error rates for the three decoding algorithms for various values of transition probabilities from the good to the bad state.N=2000,R=1/2}\label{fig:fig17}
  \end{center}
\end{figure}

\begin{figure}
\begin{center}
  \scalebox{0.3}{\includegraphics{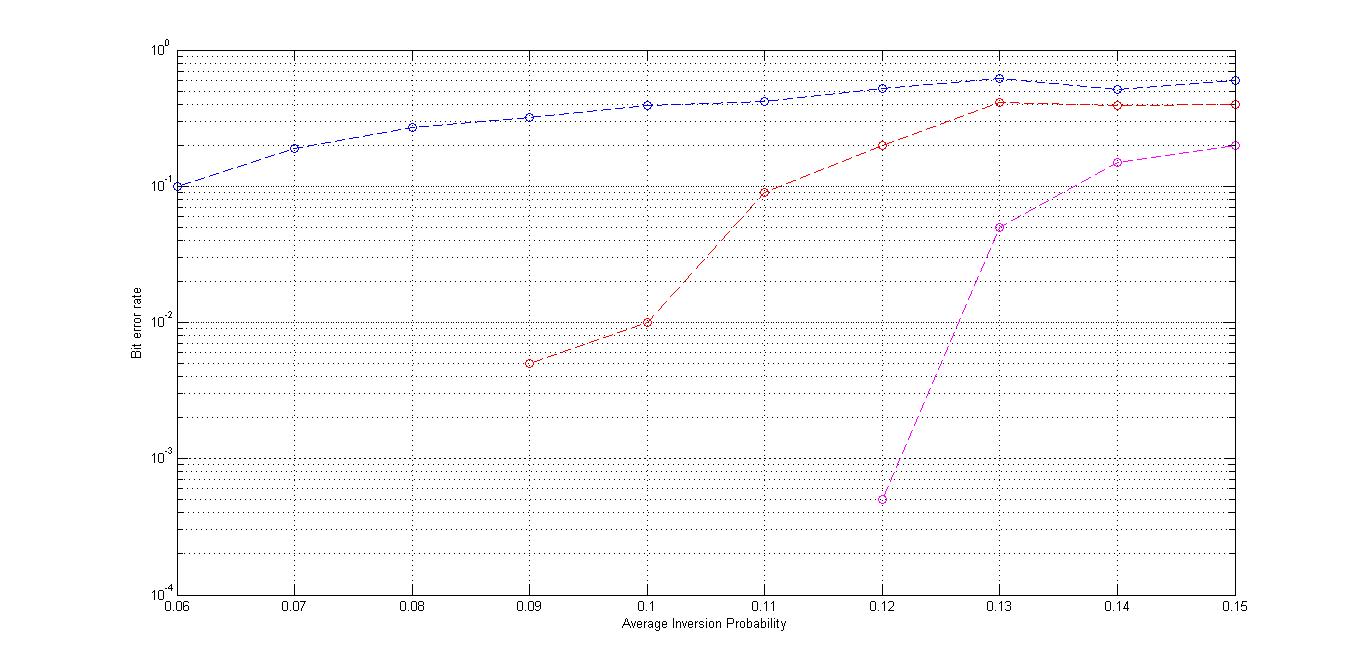}}

  \caption{Comparison of bit error rates for the three decoding algorithms for various values of average error probability of the MM channel.N=2000,R=1/2}\label{fig:fig18}
  \end{center}
\end{figure}

\begin{figure}
\begin{center}
  \scalebox{0.6}{\includegraphics{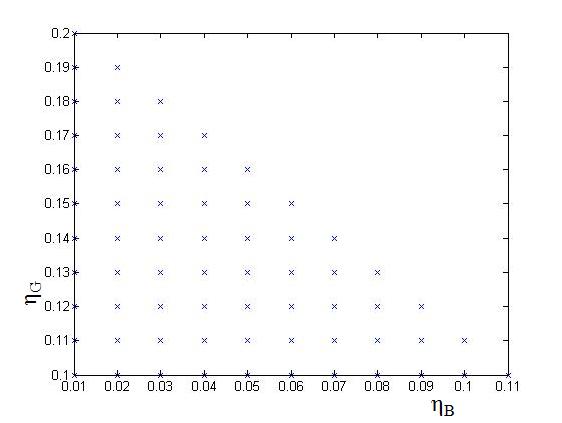}}

  \caption{Plot for a (2,4) regular rate 1/2 code showing the region of successful decoding}\label{fig:fig19}
  \end{center}
\end{figure}

\begin{figure}
\begin{center}
  \scalebox{0.6}{\includegraphics{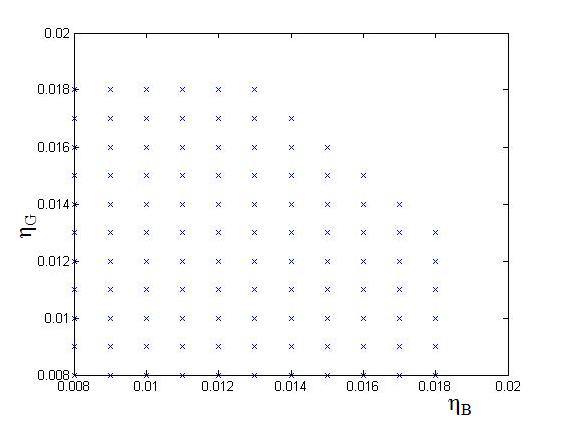}}

  \caption{Plot for a (4,6) regular rate 1/3 code showing the region of successful decoding}\label{fig:fig20}
  \end{center}
\end{figure}

\begin{enumerate}
\item Figure \ref{fig:fig10} shows comparison of bit error rates for sum-product decoder and sum-product decoder with helper block.
This is a rate $1/2$ code with $N=2000$. The two algorithms are compared for various transition probabilities.
This figure illustrates that the sum-product decoder with helper block shown in red outperforms the sum-product decoder in blue.
The sum-product decoder performance is nearly same for all transition probabilities as the average error rate is above the threshold for this algorithm.
Both the algorithms are run for $50$ iterations.
\item Figure \ref{fig:fig12} shows the comparison between same two algorithms as in figure \ref{fig:fig10} but with N=4000. These codes are designed to show excellent performances as the code length becomes large.
\item Figure \ref{fig:fig13} and figure \ref{fig:fig16} shows the convergence of the parameters of the markov model by Baum-Welch algorithm. The two figures start with different initial values for transition probabilities.
\item Figure \ref{fig:fig14} and figure \ref{fig:fig15} shows the convergence of the parameters of the markov model by Baum-Welch algorithm. The two figures start with different initial values for error probabilities in the two states.
\item Figure \ref{fig:fig17} illustrates the comparison of bit error rates for the three algorithm. The line in magenta shows the performance of Gallager's modified algorithm which scores above the other two in terms of error rate v/s the transition probability from good to bad state. This is a rate 1/2 code with blocklength = 2000.
\item Figure \ref{fig:fig18} illustrates the comparison of bit error rates for the three algorithm. The line in magenta shows the performance of Gallager's modified algorithm which scores above the other two in terms of error rate v/s the average inversion probability of the MM channel. This is a rate 1/2 code with blocklength = 2000.
\item Figure \ref{fig:fig19} displays a plot for a (2,4) regular rate 1/2 code. The stars mark the region where successful decoding can take place for the given error parameters for a two state markov model. This result also indicates that for a constant $\overline{\eta}$, increasing the difference between $\eta_{B}$ and $\eta_{G}$ leads to decrease in error probability at the decoder. Thus a larger contrast between $\eta_{B}$ and $\eta_{G}$ is better than a smaller contrast, because the states are easier to distinguish. This has been shown in \cite{a9} using technique called state scrambling.
\item Figure \ref{fig:fig20} displays a plot for a (4,6) regular rate 1/3 code. The stars mark the region where successful decoding can take place for the given error parameters for a two state markov model. Also it has been shown in \cite{a9} that increasing the length of the channel memory decreases the error probability at the decoder. This has been proved there using technique of segmentation.
\end{enumerate}

%

\chapter{Conclusion And Future Work}
In this thesis, we have discussed two new algorithms for decoding LDPC codes over channels modeled using markov modeling techniques. For some channel parameters, these algorithms seem to correct errors far beyond the random error capability of these codes with reasonable decoding complexity. Furthermore, the proposed decoding algorithms has high flexibility to
adapt channel variations because we only need to update the channel model
used in the algorithms adaptively for such variations.\\
Performance improvements are particularly noted when the contrast between the error probabilities of the two states is large. This is understandable since it is under these circumstances that the GE factor graph has the best ability to observe a state for a long period of time and differentiate between two states. Below the decoding threshold, the decoding performance improves everywhere as iterations are increased.\\
Here we have not discussed the code design problem. The performance
of an error correcting code depends greatly on the distance structure of
the coded sequences. It appears interesting research
topic to construct a class of LDPC codes matched to
Markov noise channels. This direction should be investigated in future. Many iterative techniques can be used on bipartite graphs. It would be worth investigating other schemes and comparing their expected computational complexities.\\
Recently, researchers are finding different alternatives to the sum-product algorithm to overcome the complexity issue. One of the technique that has attracted considerable attention is that based on neural networks. More hints of using this technique can be found in \cite{a10} and can be tailored for our purpose.\\
For obtaining better codes with good distance properties, differential evolution can be used as a tool for global optimization over continuous spaces. Refer \cite{a11} and \cite{a12} for the usage of this technique.


\end{document}